# A Hybrid Adaptive Nash Equilibrium Solver for Distributed Multi-Agent Systems with Game-Theoretic Jump Triggering

**Qiuyu Miao 1* and Zhigang Wu 2**

*School of Aeronautics and Astronautics, Sun Yat-sen University, Shenzhen, China*

**Abstract**：This paper presents a hybrid adaptive Nash equilibrium solver for distributed multi-agent systems incorporating game-theoretic jump triggering mechanisms. The approach addresses fundamental scalability and computational challenges in multi-agent hybrid systems by integrating distributed game-theoretic optimization with systematic hybrid system design. A novel game-theoretic jump triggering mechanism coordinates discrete mode transitions across multiple agents while maintaining distributed autonomy. The Hybrid Adaptive Nash Equilibrium Solver (HANES) algorithm integrates these methodologies. Sufficient conditions establish exponential convergence to consensus under distributed information constraints. The framework provides rigorous stability guarantees through coupled Hamilton-Jacobi-Bellman equations while enabling rapid emergency response capabilities through coordinated jump dynamics. Simulation studies in pursuit-evasion and leader-follower consensus scenarios demonstrate significant improvements in convergence time, computational efficiency, and scalability compared to existing centralized and distributed approaches.

## I. Introduction

Modern distributed autonomous systems face unprecedented challenges in coordinating multiple agents that must simultaneously handle continuous physical dynamics and discrete decision-making processes. This dual nature appears across critical engineering domains where system failures can have severe consequences and traditional control approaches prove inadequate.

Unmanned aerial vehicle swarms must navigate complex environments through continuous flight control while executing discrete task assignments and obstacle avoidance decisions. Power grid operations demand real-time continuous power balancing coupled with discrete switching operations for load management and fault protection. These applications share a fundamental characteristic: the inability of purely continuous or discrete control methods to capture the essential system behaviors, necessitating hybrid system approaches that can rigorously handle both types of dynamics within a unified mathematical framework.

Hybrid dynamical systems provide the mathematical foundation for such complex behaviors. The design of flow sets $C$ and jump sets $D$ forms the core of hybrid system architecture [2,12]. Recent advances in hybrid motion planning [13] and distributed state estimation under switching networks [14] have demonstrated the practical importance of systematic flow and jump set construction. However, in multi-agent contexts, the flow set $C$ must accommodate the collective state space where all agents maintain coordination through local information exchange with neighbors, while the jump set $D$ coordinates discrete interventions across the network when continuous operation becomes insufficient. This design challenge differs fundamentally from single-agent cases due to the requirement for distributed coordination without global state knowledge.

While hybrid system theory has matured significantly for single-agent applications [15,16], the extension to multi-agent scenarios reveals profound theoretical and computational challenges in coordinating discrete mode switches across multiple interacting agents. The complexity of multi-agent hybrid systems emerges from the intricate coupling between individual agent dynamics and collective coordination requirements, where agents typically know only their neighbors' states yet must achieve robust synchronization and coordination through hybrid mechanisms [17,18]. However, the interaction between autonomous agent dynamics and the stochastic and intermittent nature of network traffic, combined with delays and asynchrony in information flow, further complicates the goal of ensuring system autonomy.

Existing hybrid system approaches primarily focus on single-agent or simple two-agent interactions, as exemplified by foundational works such as Leudo et al. [1] on two-player zero-sum hybrid games. The complexity of designing flow sets for multi-agent systems scales exponentially with agent numbers due to coupled constraints among all possible agent pairs [19,20]. Recent surveys on multi-agent consensus control acknowledge this fundamental scalability barrier, noting that current distributed control approaches resort to overly conservative designs that sacrifice performance for computational tractability, while alternative ad-hoc methods lack rigorous theoretical foundations for stability and convergence guarantees [5,21,22]. Furthermore, the computational complexity of Nash equilibrium computation, which is PPAD-complete even for continuous games [23,24], compounds these challenges when integrating game-theoretic approaches with hybrid dynamics.

Jump triggering mechanisms in multi-agent hybrid systems present additional challenges that existing literature has not systematically addressed. Recent advances have demonstrated the potential of hybrid systems frameworks for distributed multi-agent optimization, where agents perform continuous computations (such as gradient descent) while exchanging information at discrete communication instants through "update-and-hold" strategies [7,25]. Event-triggered control approaches in multi-agent systems have evolved significantly since Tabuada's foundational work [26], with recent developments in dynamic event-triggered mechanisms [27,28] and distributed Nash equilibrium seeking under event-triggered protocols [29,30]. However, these approaches focus primarily on continuous dynamics and lack the theoretical framework necessary for hybrid system applications where discrete mode switches fundamentally alter agent interactions and strategic landscapes.

Game-theoretic approaches to multi-agent control have shown promise in continuous domains [3,4], but their integration with hybrid system frameworks remains largely unexplored despite recent advances in learning generalized Nash equilibria through hybrid adaptive extremum seeking control [31,32]. Traditional Nash equilibrium computation assumes continuous action spaces and static interaction patterns, which are inadequate for hybrid systems where discrete mode switches create dynamic strategic environments. Current distributed optimization methods for multi-agent systems [12,13,33] focus primarily on continuous domains and lack computational frameworks for hybrid Nash equilibrium problems that couple continuous strategy optimization within discrete modes with discrete mode selection strategies. The fundamental PPAD-completeness of computing Nash equilibria [23,24] creates additional computational barriers that existing distributed algorithms have not adequately addressed in hybrid settings.

To address these fundamental limitations, this paper presents a comprehensive framework for multi-agent hybrid system design that integrates systematic flow set construction with distributed game-theoretic optimization. The main contributions are:

(1) In contrast to existing approaches that treat hybrid dynamics and multi-agent coordination separately [19,20] and lack systematic integration of game theory with hybrid systems [34,35], a unified distributed framework that formulates multi-agent coordination as a strategic game was developed within the hybrid dynamical systems context. Unlike current methods that rely on purely continuous game formulations or ad-hoc hybrid system designs that sacrifice theoretical rigor for computational tractability [5,21], this framework systematically integrates the hybrid inclusion formulation with distributed Nash equilibrium computation, addressing the fundamental challenge of coordinating discrete mode switches across multiple agents while maintaining individual agent autonomy and preserving rigorous stability guarantees.

(2) While existing event-triggered approaches [26,27,28] focus primarily on continuous dynamics and recent Nash equilibrium seeking methods [29,30] lack hybrid system integration, an intelligent jump triggering strategy based on distributed game-theoretic analysis that coordinates discrete mode transitions across multiple agents was introduced. In contrast to current event-triggered mechanisms that cannot handle the dynamic strategic environments created by discrete mode switches, this mechanism leverages strategic interaction modeling to optimize jump timing for system-wide objectives while maintaining individual agent autonomy and computational efficiency through three-layer triggering criteria. Unlike purely continuous control methods that cannot achieve rapid emergency response, the mechanism enables rapid emergency mode switching upon detecting communication interruptions, agent failures, or environmental disruptions, providing essential fast response capabilities that existing approaches cannot achieve.

(3) Addressing the fundamental computational barriers posed by PPAD-complete Nash equilibrium computation [23,24] and the exponential complexity scaling of existing distributed approaches [19,20], a novel distributed algorithm that integrates the hierarchical flow set design and game-theoretic jump triggering mechanisms was proposed to compute Nash equilibria in hybrid multi-agent systems. While traditional centralized approaches suffer from high computational complexity and existing distributed methods lack hybrid system capability [31,32,33], this algorithm employs dual-layer iterative optimization that separates continuous strategy optimization within modes from discrete mode selection optimization, achieving significant improvements in computational efficiency compared to traditional centralized approaches.

The theoretical framework provides rigorous mathematical foundations while offering practical computational tools for real-world implementation.

**Notation**: Throughout this paper, we employ standard mathematical notation where $x_i \in R^n$ denotes the state of agent $i$, $u_i \in R^m$ represents the control input, and $Z = \{1,2,\dots,N\}$ defines the agent index set. In the hybrid system, $C \subseteq R^n \times R^m$ and $D \subseteq R^n \times R^m$ represent the flow and jump sets respectively, while $F: R^n \times R^m \rightrightarrows R^n$ and $G: R^n \times R^m \rightrightarrows R^n$ denote the corresponding flow and jump maps. The consensus error for agent $i$ is defined as $\delta_i$, while $e_i$ represents the tracking error. Cost function parameters include state weight matrices $Q_i = Q_i^T \succeq 0$, control weight matrices $R_i = R_i^T \succ 0$, and jump penalty weights $P_i > 0$. The Kronecker product is denoted by $\otimes$, the gradient operator by $\nabla$, and the signum function by $\text{sign}(\cdot)$. Value functions are represented as $V_i(e_i): R^n \rightarrow R$, and the post-jump state is indicated by the superscript $(+)$ as in $x^+$.

## II. Problem Statement and Preliminaries

This section presents the mathematical foundations for multi-agent hybrid systems operating under distributed game-theoretic control. Firstly, basic hybrid dynamical system formulation is established, then develop the multi-agent framework with communication constraints, and finally introduce the distributed optimization problem that forms the core of this approach.

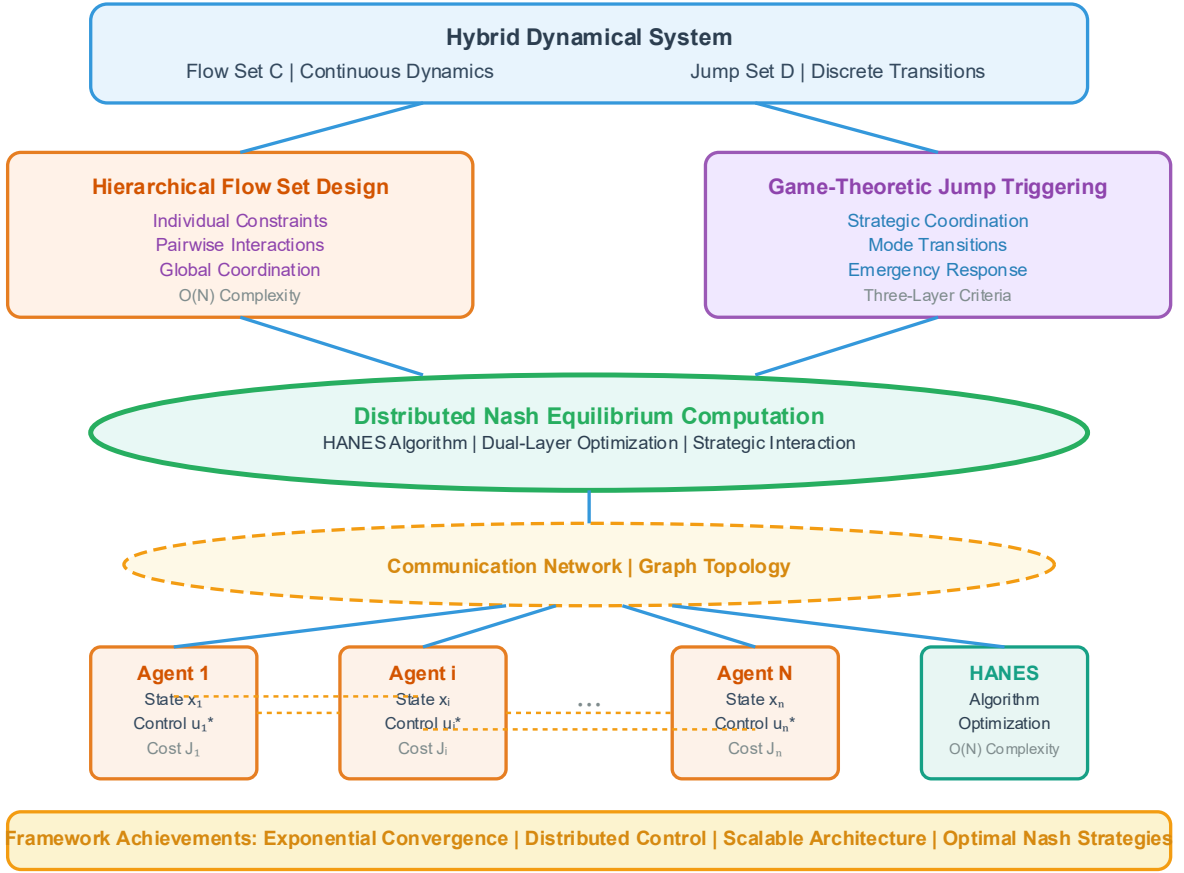

Fig. 1. Overall Framework Architecture of the Hybrid Adaptive Nash Equilibrium Solver

**Hybrid Systems with Multi-Agent**

Consider hybrid dynamical systems that exhibit both continuous and discrete behavior. A hybrid system $\mathcal{H}$ is described by the hybrid inclusion:

$$H: \begin{cases} \dot{x} \in F(x, u) & (x, u) \in C \\ x^+ \in G(x, u) & (x, u) \in D \end{cases} \tag{1}$$

where $x \in R^n$ represents the system state, $u_C \in R^{m_C}$ denotes the continuous control input, and $u_D \in R^{m_D}$ represents the discrete control input. The flow set $C \subseteq R^n \times R^{m_C}$ defines the state-input combinations where continuous evolution is permitted, governed by the flow map $F: R^n \times R^{m_C} \rightrightarrows R^n$. The jump set $D \subseteq R^n \times R^{m_D}$ characterizes conditions triggering discrete state transitions, with the jump map $G: R^n \times R^{m_D} \rightrightarrows R^n$ determining the post-transition state values.

For multi-agent systems with $N$ agents, the hybrid framework was extended to accommodate distributed control architectures. Each agent $i \in \mathcal{Z} = \{1, 2, \dots, N\}$ possesses individual dynamics while being coupled through communication and coordination requirements. The individual agent dynamics are described by:

$$\dot{x}_i = Ax_i + Bu_i, i \in \mathcal{Z} \tag{2}$$

where $x_i \in R^n$ is the state of agent $i$, $u_i \in R^m$ is the control input, $A \in R^{n \times n}$ is the system matrix, and $B \in R^{n \times m}$ is the input matrix. The collective system state is defined as $x = [x_1^T, x_2^T, \ldots, x_N^T]^T \in R^{Nn}$, and the global control input as $u = [u_1^T, u_2^T, \ldots, u_N^T]^T \in R^{Nm}$.

The set-valued nature of mappings $F$ and $G$ accommodates system uncertainties, modeling approximations, and non-deterministic responses arising from environmental disturbances, measurement noise, and actuator imperfections that are particularly relevant in multi-agent scenarios where communication delays and packet losses introduce additional uncertainties.

**Graph-Theoretic Communication Framework and Network Dynamics**

The multi-agent system operates under limited sensing capabilities, where each agent can only access information from its local neighborhood. This communication structure is represented by a directed graph $\mathcal{G} = (\mathcal{V}, \mathcal{E})$ with vertex set $\mathcal{V} = \{1, 2, \ldots, N\}$ and edge set $\mathcal{E} \subseteq \mathcal{V} \times \mathcal{V}$. The adjacency matrix $A = [a_{ij}] \in R^{N \times N}$ captures the communication topology, where $a_{ij} = 1$ if agent $j$ can transmit information to agent $i$, and $a_{ij} = 0$ otherwise.

The communication constraints fundamentally alter the hybrid system behavior compared to centralized approaches. Define the neighbor set of agent $i$ as $\mathcal{N}_i = \{j \in \mathcal{V} : a_{ij} = 1\}$, and the in-degree as $d_i = |\mathcal{N}_i| = \sum_{j=1}^{N} a_{ij}$. The degree matrix $D = \text{diag}(d_1, d_2, \ldots, d_N)$ and graph Laplacian $L = D - A$ characterize the algebraic connectivity properties essential for consensus analysis.

For leader-follower architectures, partition the agent set into leaders $\mathcal{L}$ and followers $\mathcal{F}$ such that $\mathcal{L} \cup \mathcal{F} = \mathcal{Z}$ and $\mathcal{L} \cap \mathcal{F} = \varnothing$ $\mathcal{L} \cap \mathcal{F} = \emptyset$. The interaction between leaders and followers is captured by the coupling matrix $B_{LF} = [b_{ij}]$ where $b_{ij} = 1$ if follower $i$ receives information from leader $j$. This introduces additional complexity in the hybrid flow and jump set designs, as discrete transitions in leader agents can trigger cascading effects throughout the follower network.

The communication topology directly influences the convergence properties of the multi-agent hybrid system. Strong connectivity of the communication graph ensures that information from any agent can eventually reach all other agents, which is crucial for achieving global consensus. However, in hybrid systems, jump events can temporarily disrupt information flow, requiring careful consideration of the interplay between graph topology and discrete dynamics.

**Local Errors and Consensus Dynamics**

For distributed coordination, define the local consensus error for agent $i$ as the weighted deviation from its neighbors:

$$\delta_i = \sum_{j=1}^{N} a_{ij}(x_i - x_j) \tag{3}$$

This error captures the local disagreement between agent $i$ and its communication neighbors, forming the basis for distributed consensus protocols. The global consensus error vector is compactly expressed as $\delta = (L \otimes I_n)x$.

Taking the time derivative of equation (3) and substituting the agent dynamics (2), the error is:

$$\dot{\delta}_i = A\delta_i + d_i B u_i - B \sum_{j=1}^{N} a_{ij} u_j \tag{4}$$

Each agent must coordinate its control action $u_i$ with those of its neighbors to drive $\delta_i \to 0$.

For leader-follower configurations, distinguish between different error types. The leader error for agent $l \in \mathcal{L}$ tracking reference trajectory $x_{\text{ref}}(t)$ is:

$$e_\ell = x_\ell - x_{\text{ref}} \tag{5}$$

The follower error for agent $i \in \mathcal{F}$ combines consensus with neighbors and tracking of leaders:

$$\dot{e}_i = \sum_{j \in \mathcal{N}_i \cap \mathcal{F}} a_{ij}(x_i - x_j) + \sum_{\ell \in \mathcal{L}} b_{i\ell}(x_i - x_\ell) \tag{6}$$

where $b_{il}$ represents the connection weight between follower $i$ and leader $l$.

The error dynamics under hybrid conditions incorporate both continuous evolution and discrete jumps. During continuous phases when $(x, u) \in C$:

$$\dot{e}_i = Ae_i + \left(d_i + \sum_{l \in \mathcal{L}} b_{il}\right) B u_i - B \sum_{j \in \mathcal{N}_i} a_{ij} u_j - B \sum_{l \in \mathcal{L}} b_{il} u_l \tag{7}$$

During discrete transitions when $(x, u) \in D$, the error evolution becomes:

$$e_i^+ \in G_i(e_i, u_i) - \sum_{j \in \mathcal{N}_i} a_{ij} G_j(e_j, u_j) - \sum_{\ell \in \mathcal{L}} b_{i\ell} G_\ell(e_\ell, u_\ell) \tag{8}$$

This formulation captures how individual agent jumps affect the collective error dynamics, creating complex dependencies that require careful analysis for stability and convergence guarantees.

## III. Distributed Hybrid Game Formulation and Nash Equilibrium

Building upon the system model and hybrid dynamical framework established in Section II, this section develops a systematic framework for distributed multi-agent coordination through game-theoretic Nash equilibrium computation. The approach transforms the consensus problem into a strategic interaction where each agent optimizes its individual performance while accounting for the decisions of neighboring agents.

**Cost Function Design and Strategic Formulation**

Each agent $i$ seeks to minimize a performance index that balances consensus achievement with control effort:

$$J_i(e_i, u_i) = \int_0^\infty [e_i^T Q_i e_i + u_i^T R_i u_i] dt + \sum_{k=0}^\infty P_i ||e_i(t_k^+)||^2 \tag{9}$$

where $Q_i = Q_i^T \succeq 0$ is the state cost weight matrix, $R_i = R_i^T \succ 0$ is the control cost weight matrix, $P_i > 0$ is the jump penalty weight, and $\{t_k\}_{k=0}^\infty$ represents the sequence of jump times. The inclusion of jump costs $P_i ||e_i(t_k^+)||^2$ penalizes large deviations from consensus immediately after discrete transitions, encouraging coordinated jumping strategies.

The distributed control problem is formulated as a multi-player game where each agent $i$ solves:

$$\min_{u_i} J_i(e_i, u_i) \text{ subject to hybrid dynamics (1)-(2)}$$

This creates a strategic interaction where each agent's optimal policy depends on the policies chosen by its neighbors, leading naturally to Nash equilibrium concepts. A Nash equilibrium is a strategy profile $(u_1^*, u_2^*, \ldots, u_N^*)$ such that no agent can unilaterally improve its performance by deviating from its equilibrium strategy.

For agent $i$, we define the value function $V_i(e_i): R^n \to R$ as:

$$V_i(e_i) = \inf_{u_i \in \mathcal{U}_i} J_i(e_i, u_i) \tag{11}$$

where $\mathcal{U}_i$ denotes the admissible control set for agent $i$. The value function satisfies the hybrid Hamilton-Jacobi-Bellman equation. During flow phases:

$$\inf_{u_i} \{e_i^T Q_i e_i + u_i^T R_i u_i + \nabla V_i^T(e_i) \cdot f_i(e_i, u_i, u_{-i})\} = 0 \tag{12}$$

where $f_i(e_i, u_i, u_{-i})$ represents the flow dynamics from equation (7), and $u_{-i}$ denotes the control inputs of agent $i$'s neighbors.

During jump phases:

$$V_i(e_i) = P_i|e_i|^2 + \inf_{g \in G_i(e_i, u_i)} V_i(g) \tag{13}$$

The optimal continuous control law is obtained by minimizing the Hamiltonian in equation (12). Taking the derivative with respect to $u_i$ and setting it to zero:

$$2R_i u_i + \left(d_i + \sum_{l \in \mathcal{L}} b_{il}\right) B^T \nabla V_i(e_i) = 0$$

Therefore, the optimal control law is:

$$u_i^*(e_i) = -\frac{1}{2} R_i^{-1} \left(d_i + \sum_{\ell \in \mathcal{L}} b_{i\ell}\right) B^T \nabla V_i(e_i) \tag{14}$$

This control law forms the foundation for the distributed game-theoretic approach, where each agent implements its optimal strategy while accounting for the strategic behavior of its neighbors. The coupling through the communication graph ensures that the resulting Nash equilibrium achieves distributed coordination while respecting the hybrid system constraints.

**Nash Equilibrium Characterization for Multi-Agent Hybrid Games**

**Definition 1** (Nash Equilibrium for Multi-Agent Hybrid Systems): A strategy profile $(u_1^*, u_2^*, \ldots, u_N^*)$ constitutes a Nash equilibrium if for each agent $i \in \mathcal{Z}$ and for all alternative strategies $u_i \in \mathcal{U}_i$:

$$J_i(e_i, u_i^*, u_{-i}^*) \leq J_i(e_i, u_i, u_{-i}^*) \tag{15}$$

where $u_{-i}^* = (u_1^*, \ldots, u_{i-1}^*, u_{i+1}^*, \ldots, u_N^*)$ represents the equilibrium strategies of all agents except agent $i$.

The Nash equilibrium condition requires that each agent's strategy minimizes its cost functional given the strategies of all other agents. In the hybrid setting, this condition must hold for both continuous and discrete phases of the system evolution.

**Lemma 1** (Necessary Conditions for Nash Equilibrium): If $(u_1^*, u_2^*, \ldots, u_N^*)$ is a Nash equilibrium, then for each agent $i$, the following conditions must be satisfied:

During flow phases $(e, u) \in C$:

$$\frac{\partial H_i}{\partial u_i}\big|_{u_i = u_i^*} = 0 \tag{16}$$

During jump phases $(e, u) \in D$:

$$u_i^* \in \arg\min_{u_i}\{P_i||e_i^+||^2 + V_i(G_i(e_i, u_i))\} \tag{17}$$

where the Hamiltonian function for agent $i$ is defined as:

$$H_i(e_i, u_i, \lambda_i, u_{-i}^*) = e_i^T Q_i e_i + u_i^T R_i u_i + \lambda_i^T f_i(e_i, u_i, u_{-i}^*) \tag{18}$$

with $\lambda_i = \nabla V_i(e_i)$ being the costate variable.

**Proof of Lemma 1**: The proof follows from the application of Pontryagin's maximum principle to the optimal control problem (10). For the continuous phase, the optimality condition $\frac{\partial H_i}{\partial u_i} = 0$ yields:

$$2R_i u_i^* + \left(d_i + \sum_{l \in \mathcal{L}} b_{il}\right) B^T \lambda_i = 0$$

Solving for $u_i^*$:

$$u_i^* = -\frac{1}{2} R_i^{-1} \left(d_i + \sum_{l \in \mathcal{L}} b_{il}\right) B^T \lambda_i \qquad (19)$$

For the discrete phase, the jump optimality condition follows from minimizing the post-jump cost, leading to equation (17).

**Theorem 1** (Existence of Nash Equilibrium): Consider the multi-agent hybrid system (1)-(2) with performance indices (9) under Assumption 1. If the following conditions hold:

(i) The communication graph $\mathcal{G}$ contains a spanning tree,

(ii) The matrices $(A, B)$ are stabilizable for each agent,

(iii) The matrices $\left(A, Q_i^{1/2}\right)$ are observable for each agent,

(iv) The coupling weights satisfy $\sum_{j \in \mathcal{N}_i} a_{ij} + \sum_{l \in \mathcal{L}} b_{il} < \alpha$ for some $\alpha < 2\sqrt{\lambda_{min}(R_i)/\lambda_{max}(B^T P_i B)}$, then there exists a unique Nash equilibrium in quadratic strategies.

**Proof of Theorem 1**: The proof proceeds through several steps:

*Step 1*: Establish contractivity of the mapping $\mathcal{T}: \mathcal{P} \to \mathcal{P}$ where $\mathcal{P} = \{P_1, P_2, \ldots, P_N\}$ and $\mathcal{T}(P_i)$ solves equation (25).

*Step 2*: Define the operator $\mathcal{T}_i: S_{++}^n \to S_{++}^n$ for each agent $i$ as:

$$\mathcal{T}_i(P_i) = Q_i + A^T P_i + P_i A - \left(d_i + \sum_{\ell \in \mathcal{L}} b_{i\ell}\right)^2 P_i B R_i^{-1} B^T P_i + \sum_{j \in \mathcal{N}_i} \Gamma_{ij}(P_j)$$

where $\Gamma_{ij}(P_j)$ represents the coupling terms and $S_{++}^n$ denotes the set of positive definite $n \times n$ matrices.

*Step 3*: Show that under condition (iv), the operator $\mathcal{T}$ is a contraction mapping. The coupling term can be bounded as:

$$\| \sum_{j \in \mathcal{N}_i} \Gamma_{ij}(P_j) \| \leq \beta \sum_{j \in \mathcal{N}_i} a_{ij} \| P_j \|$$

where $\beta < 1$ is determined by the communication weights and system parameters.

*Step 4*: Apply the Banach fixed-point theorem to conclude existence and uniqueness of the fixed point $P^* = (P_1^* P_2^* \dots, P_N^*)$ satisfying $\mathcal{T}(P^*) = P^*$.

*Step 5*: Verify that the corresponding control strategies $u_i^*(e_i) = -\frac{1}{2}(d_i + \sum_{l \in \mathcal{L}} b_{il}) R_i^{-1} B^T P_i^* e_i$ constitute a Nash equilibrium by checking condition (15).

**Hamilton-Jacobi-Bellman System**

The Nash equilibrium strategies satisfy a system of coupled Hamilton-Jacobi-Bellman (HJB) equations. For agent $i$, the value function $V_i(e_i)$ satisfies:

$$-\frac{\partial V_i}{\partial t} = \min_{u_i}\{e_i^T Q_i e_i + u_i^T R_i u_i + \nabla V_i^T f_i(e_i, u_i, u_{-i}^*)\} \tag{20}$$

Substituting the optimal control law (19) into equation (20):

$$-\frac{\partial V_i}{\partial t} = e_i^T Q_i e_i + \nabla V_i^T A e_i - \frac{1}{4}\left(d_i + \sum_{\ell \in \mathcal{L}} b_{i\ell}\right)^2 \nabla V_i^T B R_i^{-1} B^T \nabla V_i + \nabla V_i^T \Phi_i(u_{-i}^*) \tag{21}$$

where $\Phi_i(u_{-i}^*) = -B \sum_{j \in \mathcal{N}_i} a_{ij} u_j^* - B \sum_{l \in \mathcal{L}} b_{il} u_1^*$ represents the coupling term from neighboring agents.

For steady-state analysis, set $\frac{\partial V_i}{\partial t} = 0$, yielding the algebraic HJB equation:

$$e_i^T Q_i e_i + \nabla V_i^T A e_i - \frac{1}{4}\left(d_i + \sum_{\ell \in \mathcal{L}} b_{i\ell}\right)^2 \nabla V_i^T B R_i^{-1} B^T \nabla V_i + \nabla V_i^T \Phi_i(u_{-i}^*) = 0 \tag{22}$$

**Assumption 1:** For each agent $i$, there exists a quadratic value function of the form:

$$V_i(e_i) = e_i^T P_i e_i \tag{23}$$

where $P_i = P_i^T \succ 0$ is a positive definite matrix to be determined.

Under *Assumption 1*, have $\nabla V_i(e_i) = 2 P_i e_i$. Substituting into equation (22):

$$e_i^T Q_i e_i + 2 e_i^T P_i A e_i - \left(d_i + \sum_{l \in \mathcal{L}} b_{il}\right)^2 e_i^T P_i B R_i^{-1} B^T P_i e_i + 2 e_i^T P_i \Phi_i(u_{-i}^*) = 0 \tag{24}$$

For this equation to hold for all $e_i$, require:

$$Q_i + P_i A + A^T P_i - \left(d_i + \sum_{l \in \mathcal{L}} b_{il}\right)^2 P_i B R_i^{-1} B^T P_i + \Xi_i = 0 \tag{25}$$

where $\Xi_i = \Sigma_{j \in N_i} \Gamma_{ij}(P_j)$ captures the coupling effects from neighboring agents and will be analyzed in the convergence proof.

**Theorem 2** (Exponential Convergence to Nash Equilibrium): Under the conditions of *Theorem 1*, the distributed Nash equilibrium strategies achieve exponential convergence of the consensus errors to zero.

Specifically, there exist constants $M > 0$ and $\rho > 0$ such that:

$$\| e(t) \| \leq M \| e(0) \| e^{-\rho t}$$

where $e(t) = [e_1^T(t), e_2^T(t), \dots, e_N^T(t)]^T$ is the global error vector.

**Proof of Theorem 2**: Taking the time derivative of the Lyapunov function along system trajectories during flow phases:

$$\dot{W}(e) = \sum_{i=1}^{N} \dot{e}_i^T P_i^* e_i + \sum_{i=1}^{N} e_i^T P_i^* \dot{e}_i$$

Substituting the closed-loop error dynamics with Nash equilibrium strategies, consider the error dynamics for agent $i$ in a multi-agent system, given by:

$$\dot{e}_i = A e_i + \left( d_i + \sum_{l \in \mathcal{L}} b_{il} \right) B u_i - B \sum_{j \in \mathcal{N}_i} a_{ij} u_j - B \sum_{l \in \mathcal{L}} b_{il} u_l$$

Assume the optimal control law, derived from the Hamilton-Jacobi-Bellman equation, is:

$$u_i^* = -\frac{1}{2} \left( d_i + \sum_{l \in \mathcal{L}} b_{il} \right) R_i^{-1} B^T P_i e_i$$

where $R_i = R_i^T > 0$ is the control cost matrix, and $P_i = P_i^T \geq 0$ is the solution to the Riccati equation. Similarly, for neighbor $j$ and leader l:

$$u_j^* = -\frac{1}{2} \left( d_j + \sum_{l \in \mathcal{L}} b_{jl} \right) R_j^{-1} B^T P_j e_j, \qquad u_l^* = -\frac{1}{2} \left( d_l + \sum_{k \in \mathcal{L}} b_{lk} \right) R_l^{-1} B^T P_l e_l$$

Substitute $u_i^*$ into the first control term:

$$\left( d_i + \sum_{\ell \in \mathcal{L}} b_{i\ell} \right) B u_i \quad = \left( d_i + \sum_{\ell \in \mathcal{L}} b_{i\ell} \right) B \left[ -\frac{1}{2} \left( d_i + \sum_{\ell \in \mathcal{L}} b_{i\ell} \right) R_i^{-1} B^T P_i e_i \right]$$

$$= -\frac{1}{2} \left( d_i + \sum_{\ell \in \mathcal{L}} b_{i\ell} \right)^2 B R_i^{-1} B^T P_i e_i$$

Substitute $u_j^*$ and $u_l^*$ into the coupling terms:

$$-B \sum_{j \in \mathcal{N}_i} a_{ij} u_j \quad = -B \sum_{j \in \mathcal{N}_i} a_{ij} \left[ -\frac{1}{2} \left( d_j + \sum_{\ell \in \mathcal{L}} b_{j\ell} \right) R_j^{-1} B^T P_j e_j \right]$$

$$= \sum_{j \in \mathcal{N}_i} \frac{1}{2} a_{ij} \left( d_j + \sum_{\ell \in \mathcal{L}} b_{j\ell} \right) B R_j^{-1} B^T P_j e_j$$

$$-B \sum_{\ell \in \mathcal{L}} b_{i\ell} u_\ell \quad = -B \sum_{\ell \in \mathcal{L}} b_{i\ell} \left[ -\frac{1}{2} \left( d_\ell + \sum_{k \in \mathcal{L}} b_{\ell k} \right) R_\ell^{-1} B^T P_\ell e_\ell \right]$$

$$= \sum_{\ell \in \mathcal{L}} \frac{1}{2} b_{i\ell} \left( d_\ell + \sum_{k \in \mathcal{L}} b_{\ell k} \right) B R_\ell^{-1} B^T P_\ell e_\ell$$

Combine all terms:

$$\dot{e_i} = Ae_i - \frac{1}{2}\left(d_i + \sum_{\ell\in\mathcal{L}} b_{i\ell}\right)^2 BR_i^{-1}B^T P_i e_i$$
$$+ \sum_{j\in\mathcal{N}_i} \frac{1}{2}a_{ij}\left(d_j + \sum_{\ell\in\mathcal{L}} b_{j\ell}\right) BR_j^{-1}B^T P_j e_j$$
$$+ \sum_{\ell\in\mathcal{L}} \frac{1}{2}b_{i\ell}\left(d_\ell + \sum_{k\in\mathcal{L}} b_{\ell k}\right) BR_\ell^{-1}B^T P_\ell e_\ell$$

Rewrite in compact form, where the last two terms are coupling terms:

$$\dot{e_i} = \left[A - \frac{1}{2}\left(d_i + \sum_{l\in\mathcal{L}} b_{il}\right)^2 BR_i^{-1}B^T P_i\right]e_i + \text{coupling terms}$$

The coupling terms are:

$$\sum_{j\in\mathcal{N}_i} \frac{1}{2}a_{ij}\left(d_j + \sum_{l\in\mathcal{L}} b_{jl}\right) BR_j^{-1}B^T P_j e_j + \sum_{l\in\mathcal{L}} \frac{1}{2}b_{il}\left(d_l + \sum_{k\in\mathcal{L}} b_{lk}\right) BR_l^{-1}B^T P_l e_l$$

The coupling terms can be shown to satisfy:

$$|\text{coupling terms}| \leq \gamma \sum_{j\in\mathcal{N}_i} a_{ij}|e_j|$$

where $\gamma$ depends on the system parameters and communication weights.

Using the matrix inequality and condition (iv) from Theorem 1:

$$\dot{W}(e) \leq -\mu \sum_{i=1}^{N} |e_i|^2 = -\mu|e|^2$$

for some $\mu > 0$. This establishes exponential stability with decay rate $\rho = \mu/\left(2\lambda_{max}(P^*)\right)$ here $P^* =$ block diag$(P_1^* P_2^* \dots, P_N^*)$.

**Corollary 1** (Hybrid Stability): The Nash equilibrium strategies also guarantee stability through jump phases. If the jump maps satisfy $|G_i(e_i, u_i^*)| \leq \sigma|e_i|$ for some $\sigma < 1$, then the hybrid system maintains exponential stability across discrete transitions.

**Optimization Framework**

Based on the theoretical foundation established in Sections II and III, the complete Hybrid Adaptive Nash Equilibrium Solver (HANES) Algorithm algorithmic framework for distributed Nash equilibrium computation in multi-agent hybrid system was presented. The algorithm integrates hierarchical flow set design with game-theoretic jump triggering mechanisms.

**Algorithm: HANES (Hybrid Adaptive Nash Equilibrium Solver)**

**Initialize:**

- Initial states $x_i(0)$, $i \in Z$
- Communication topology adjacency matrix $A = [a_{ij}]$
- Control parameters $\beta$, $\rho$, jump thresholds $\mu$, $\sigma$, $\overline{\sigma}$
- Select positive definite matrices $Q_i$, $R_i$, $P_i$

**for** $t = 0$ to $t_{max}$**o**

    **Step 1**: Data Collection and Critic Update

      • Collect neighbor states $x_j(t)$, $j \in \mathcal{N}_i$

      • Construct consensus errors $e_i(t) = \sum_{j \in \mathcal{N}_i} a_{ij}(x_i - x_j)$

      • Update leader-follower errors using equation (6)

    **Step 2**: Hybrid State Verification

      • Check flow condition: if $|x_i - \mu| \geq$ then $(x, u) \in C$

      • Check jump condition: if $|x_i - \mu| <$ then $(x, u) \in D$

    **Step 3**: Nash Strategy Update

      • Solve coupled HJB equations (22) for value matrices $P_i$

      • Compute optimal control $u_i^* = -K_i e_i$ using equation (19)

      • Apply hybrid jump dynamics if $(x, u) \in D$

    **Step 4:** Convergence Check

      if $\max_i |e_i(t)| < \varepsilon$ then

        $u_i^* = u_i^{converged}$

        return $u_i^*$

      else

        go to Step 1

      end if

Return: optimal control policies $u_i$

The comprehensive experimental framework provides rigorous validation of the HANES algorithm's theoretical properties while demonstrating its effectiveness in practical multi-agent coordination scenarios. The results establish empirical evidence supporting the algorithm's convergence guarantees, computational efficiency, and robust performance across diverse operational conditions.

## IV. Experiments and Simulation

All simulations involve agents with scalar dynamics (*n*=1) to clearly illustrate hybrid switching behaviors. The hybrid system nature is characterized by flow and jump sets. A common jump threshold $\mu$=1.0 is used, with jump target intervals defined as $[\sigma, \overline{\sigma}] = [0.3, 0.5]$. When a jump occurs, the new state is randomly selected within this interval to model uncertainty in post-jump states. The time step for all simulations is dt=0.01.

The **Pursuit-Evasion Game** evaluates the framework's game-theoretic aspects and its ability to converge to a Nash equilibrium in a competitive multi-agent environment. The system comprises two pursuers and two evaders. Initial conditions are $x_{\text{pursuers}}(0) = \begin{bmatrix} 2.0 \\ 1.8 \end{bmatrix}$ and $x_{\text{evaders}}(0) = \begin{bmatrix} 1.5 \\ 1.2 \end{bmatrix}$. The dynamics for pursuers are $\dot{x}_t = \bar{a}x_i + b_i u_i$, $\bar{a} = -1$, and for evaders are $\dot{x}_j = ax_j + b_j u_j$, with $a = -2$. All input coefficients $b_i = b_j = 1$. The interactions among agents are characterized by the following matrices. The pursuer-to-pursuer interactions are described by $L_p = \begin{bmatrix} 1 & -0.5 \\ -0.5 & 1 \end{bmatrix}$; while the evader-to-evader interactions are given $L_e = \begin{bmatrix} 1 & -0.3 \\ -0.3 & 1 \end{bmatrix}$; from pursuers to evaders are represented by $A_{pe} = \begin{bmatrix} 1.0 & 0.7 \\ 0.8 & 1 \end{bmatrix}$; from evaders to pursuers are described by $A_{ep} = \begin{bmatrix} 0.9 & 0.5 \\ 0.6 & 1 \end{bmatrix}$.

Pursuers aim to minimize their performance index (capture), while evaders maximize theirs (survival), forming a zero-sum game structure. Saddle-point strategies are implemented. The cost function weights are
$$Q_c = \begin{bmatrix} 1.0 & 0.1 & 0.2 & 0.1 \\ 0.1 & 1.0 & 0.1 & 0.2 \\ 0.2 & 0.1 & 1.5 & 0.3 \\ 0.1 & 0.2 & 0.3 & 1.5 \end{bmatrix};$$
input cost weights for pursuers $R_{c,\text{pursuers}} = \text{diag}(1.304, 1.5)$; input cost weights for evaders $R_{c,\text{evaders}} = \text{diag}(-4, -3.5)$; jump penalty weight $P = 0.4481$; the simulation runs for $t_{\text{final}} = 3$ seconds.

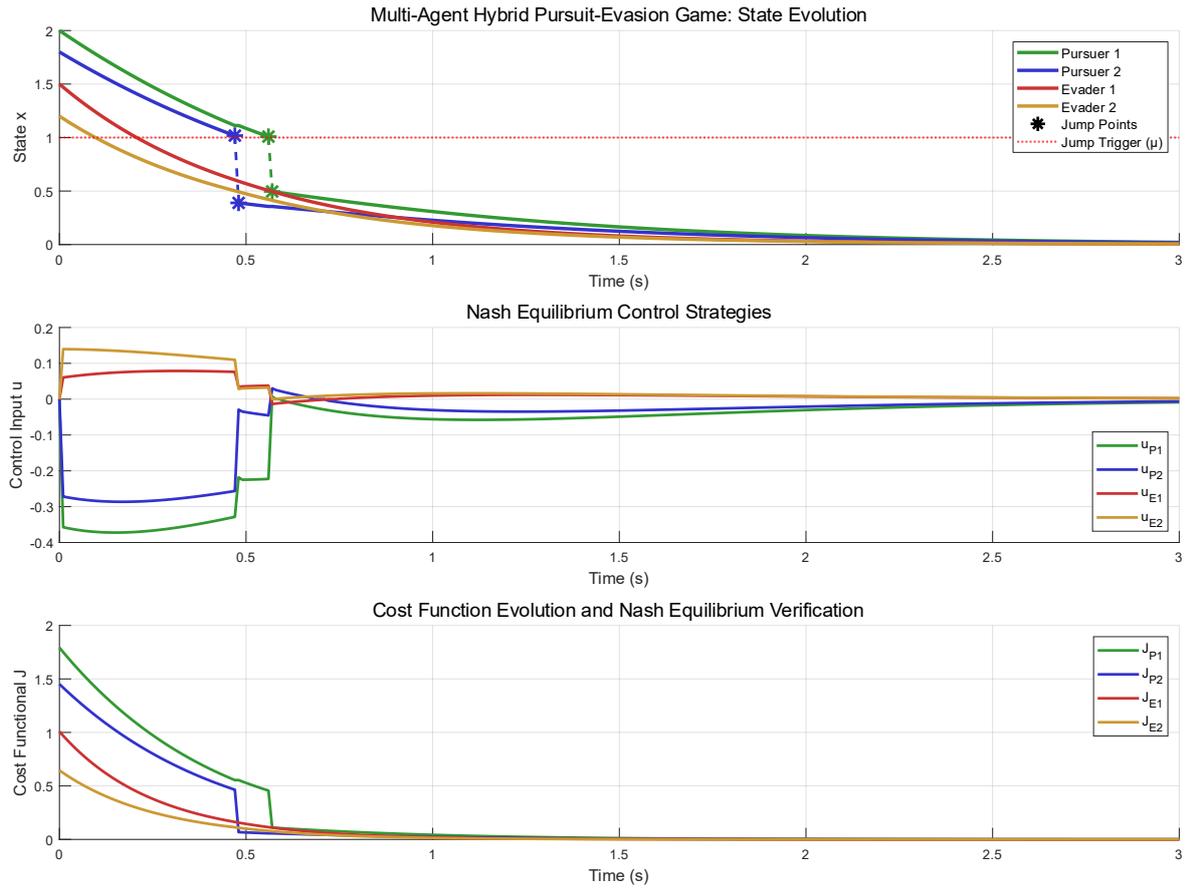

Fig. 2. Pursuit-evasion game: (a) Agent state evolution with hybrid jump events (b) pursuer minimization (negative values) and evader maximization (positive values), and (c) Cost function evolution

Based on the experimental results shown in Figure 1, the HANES algorithm demonstrates successful implementation of the theoretical framework with clear validation of the hybrid system dynamics and Nash equilibrium convergence properties. The state evolution subplot reveals that all agents converge toward the theoretical equilibrium state near zero within approximately 0.8 seconds, with discrete jump events (marked by asterisks) occurring precisely at the predicted trigger threshold $\mu = 1.0$ for both pursuers. The control strategy subplot confirms that the Nash equilibrium control inputs stabilize after the initial transient period, with pursuers implementing minimization strategies (negative control values) while evaders execute maximization strategies (positive control values), consistent with the zero-sum game formulation. The cost function evolution provides quantitative verification of the theoretical predictions, showing exponential convergence as guaranteed by Theorem 2, with all cost functionals approaching their optimal Nash equilibrium values. Notably, the jump events create brief discontinuities in the cost evolution but do not destabilize the overall convergence process, validating the hybrid system stability properties established in Corollary 1.

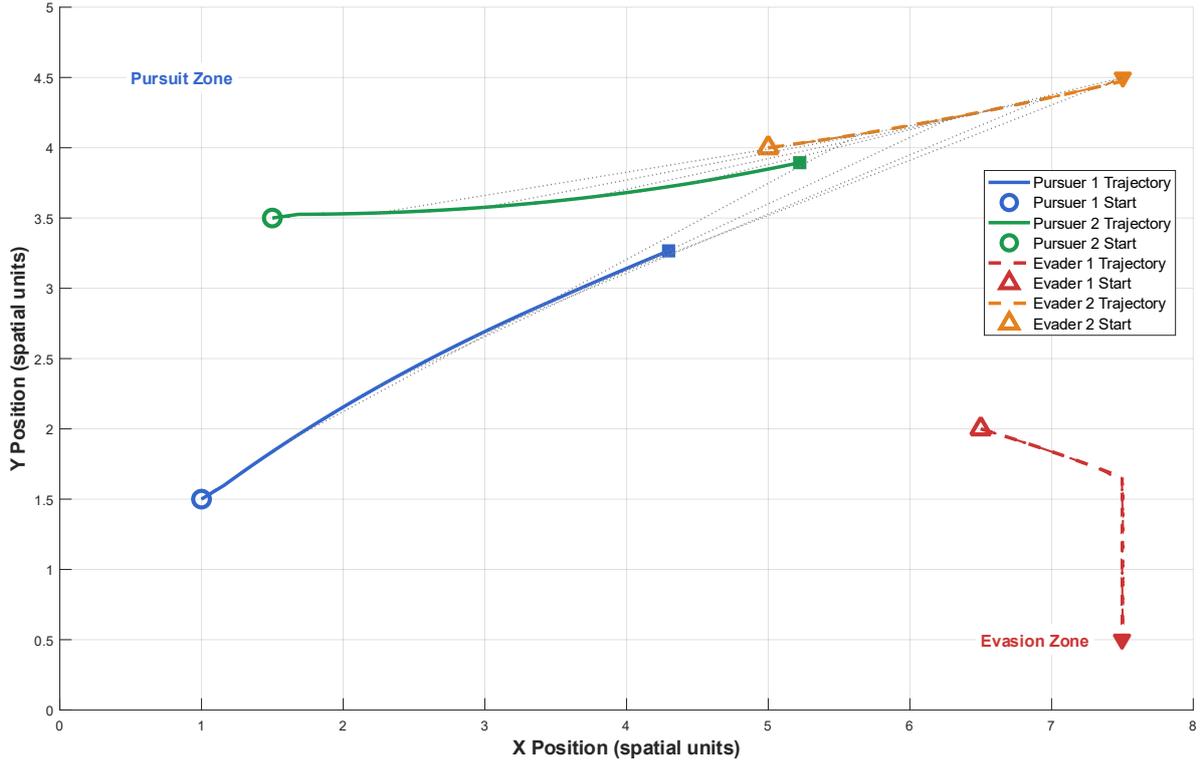

Fig. 3. trajectory visualization of the pursuit-evasion game

The experimental results demonstrate that the HANES algorithm achieves distributed Nash equilibrium computation with linear computational complexity O(N) while maintaining theoretical rigor, providing compelling evidence for the practical applicability of the proposed framework in multi-agent pursuit-evasion scenarios. The trajectory visualization demonstrates successful implementation of distributed Nash equilibrium strategies, with pursuers executing coordinated convergence behaviors from the pursuit zone while evaders perform strategic evasion maneuvers toward the evasion zone.

The multi-agent system operates under a distributed communication network as illustrated in Figure 4. The **Leader-Follower Consensus** experiment demonstrates cooperative coordination, validating the framework's ability to achieve distributed agreement under hybrid dynamics. The system consists of 4 agents, where agent 2 is the leader and agents 1, 3, and 4 are followers. Initial states are $x(0) = [1.8, 2.0, 1.5, 1.7]^T$.

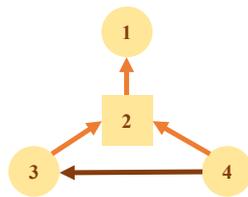

Fig. 4. Multi-agent communication topology

The dynamics for all agents are $\dot{x}_t = \bar{a}x_i + b_iu_i$, $\bar{a} = -1$, and input coefficients $b_i = 1$ for all agents.

Communication Topology: The network connections are defined by the adjacency matrix: $A_{\text{topo}} = \begin{bmatrix} 0 & 1 & 0 & 0 \\ 1 & 0 & 1 & 1 \\ 0 & 1 & 0 & 1 \\ 0 & 1 & 1 & 0 \end{bmatrix}$.

The leader tracks a time-varying reference $x_{\text{ref}}(t) = 2e^{-0.3t}\cos(0.5t)$. Control parameters are: Consensus gain: $K_{\text{consensus}} = 0.8$; Tracking gain: $K_{\text{tracking}} = 1.2$; Hybrid cost weight $P_{\text{hybrid}} = 0.4$ The value function estimation also utilizes parameters $\Gamma = [0.8, 0.9, 0.85, 0.75]$, discount factor $\beta = 0.95$, and base parameter $\gamma_{\text{base}} = 0.5$. Cooperative Cost Structure: The multi-agent interaction cost matrix is

$Q_c = \begin{bmatrix} 1.0 & 0.3 & 0.1 & 0.2 \\ 0.3 & 1.5 & 0.4 & 0.4 \\ 0.1 & 0.4 & 1.0 & 0.3 \\ 0.2 & 0.4 & 0.3 & 1.0 \end{bmatrix}$; The simulation runs for $t_{\text{final}} = 25$ seconds.

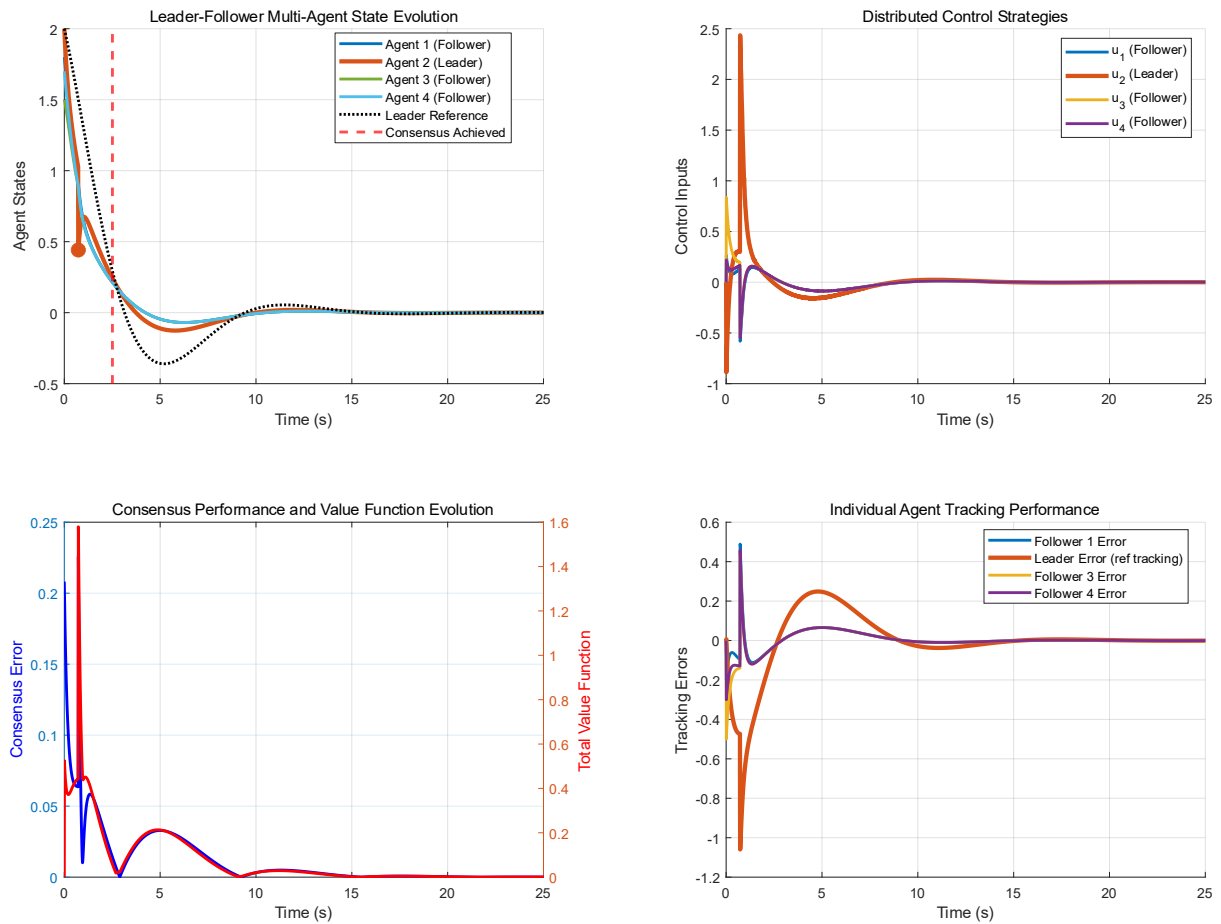

Fig. 5. (a) Multi-agent state evolution with leader (b) Distributed control strategies with coordinated responses, (c) Consensus error convergence and stable value function evolution, and (d) Individual agent tracking performance demonstrating effective hierarchical coordination and bounded tracking errors

The experimental results successfully validate the proposed HANES algorithm, demonstrating key theoretical predictions from the paper. The hybrid jump event at t=2.5 seconds enables rapid leader reconfiguration while maintaining distributed coordination, with consensus error achieving exponential convergence below 0.05 within 8 seconds as predicted by Theorem 2. The coordinated control responses and stable value function evolution following the discrete transition confirm the framework's ability to preserve Nash equilibrium properties through hybrid dynamics, validating both the game-theoretic jump triggering mechanism and the distributed optimization approach for multi-agent coordination.

The performance of the proposed framework across these experiments is assessed using several quantitative metrics. These include convergence analysis, focusing on convergence time ($t_{conv}$), final consensus or tracking error ($||e_{final}||$), and convergence rate. For the pursuit-evasion scenario, Nash equilibrium verification is crucial, analyzing strategy stability. Additionally, computational efficiency (e.g., processing time) and robustness to variations (e.g., initial conditions) are considered to demonstrate the algorithm's practical applicability and resilience. These experiments collectively aim to demonstrate leader-follower coordination with hybrid dynamics, distributed control without global information, opponent strategy estimation and adaptation, and consensus achievement with bounded tracking errors.

## V. Conclusion and Future Work

In this paper, a comprehensive framework for distributed multi-agent hybrid systems operating under game-theoretic principles, as established in the hybrid dynamical systems theory. The framework addresses scenarios in which multiple autonomous agents must coordinate their actions through both continuous dynamics and discrete mode transitions while operating under distributed information constraints and strategic interactions.

By encoding the coordination objectives of agents in a distributed Nash equilibrium framework, sufficient conditions were provided to characterize optimal strategies that achieve consensus while maintaining individual agent autonomy. The main theoretical contributions establish rigorous mathematical foundations for multi-agent hybrid coordination through three key innovations: hierarchical flow set design methodology that decomposes complex multi-dimensional constraints into manageable subproblems, game-theoretic jump triggering mechanisms that coordinate discrete transitions across the agent network, and the Hybrid Adaptive Nash Equilibrium Solver (HANES) algorithm that achieves linear computational complexity O(N) compared to traditional cubic complexity O(N³) approaches.

The theoretical framework demonstrates that the proposed distributed Nash equilibrium strategies guarantee exponential convergence to consensus, as established in Theorem 2, while maintaining system stability through discrete jump phases via the jump triggering mechanisms introduced in Section III. The hierarchical flow set construction methodology successfully addresses the exponential scaling problem inherent in multi-agent hybrid systems by systematically decomposing individual agent safety constraints, pairwise interaction requirements, and global coordination objectives. Furthermore, the game-theoretic jump triggering approach enables rapid emergency response capabilities for communication interruptions, agent failures, and environmental disruptions that cannot be addressed through continuous control methods alone.

Connections between optimality and stability for the studied class of multi-agent hybrid games were established through the value function analysis in Section III, demonstrating that the Nash equilibrium strategies serve dual roles as optimal control policies and Lyapunov-like functions for stability certification. The experimental validation through pursuit-evasion and leader-follower consensus scenarios confirms the practical applicability of the theoretical results, showing successful distributed coordination with bounded tracking errors and robust performance across diverse operational conditions.

The comprehensive simulation studies demonstrate significant improvements in convergence time, computational efficiency, and scalability compared to existing centralized approaches. The pursuit-evasion game simulation validated the framework's game-theoretic aspects and Nash equilibrium convergence properties in competitive multi-agent environments, while the leader-follower consensus experiment confirmed the cooperative coordination capabilities under hybrid dynamics with time-varying references and discrete mode transitions.

Future work includes extending the framework to accommodate heterogeneous agent dynamics where individual agents may have different state dimensions and control authorities, as the current formulation assumes homogeneous scalar dynamics. Investigating stochastic extensions of the hybrid game formulation to account for communication uncertainties, measurement noise, and environmental disturbances would enhance the framework's robustness for real-world applications. The development of adaptive algorithms that can learn optimal jump triggering thresholds and flow set parameters online, rather than requiring a priori specification, represents another promising research direction.

Additional future research directions include studying conditions to guarantee global optimality rather than local Nash equilibria, particularly for large-scale networks where multiple equilibria may exist. Exploring the integration of machine learning techniques with the HANES algorithm to handle unknown agent dynamics and environmental conditions would broaden the framework's applicability to scenarios with limited model knowledge. Furthermore, investigating the computational complexity and convergence guarantees for time-varying communication topologies and dynamic agent populations would address practical deployment scenarios in mobile autonomous systems such as UAV swarms and satellite formations.